\documentclass{elsart}

\usepackage{graphicx}               

\begin{document}

\begin{frontmatter}

\title{Heavy Residue Formation in 20 MeV/nucleon $^{197}$Au + $^{90}$Zr
       Collisions}

\author[osu]{G.A. Souliotis}$^{,1}$ \thanks{\footnotesize  
             Corresponding author: soulioti@comp.tamu.edu.
             Current Address: Cyclotron Institute,
             Texas A\&M Univ., College Station, TX 77843.},
\author[osu]{W. Loveland},
\author[ucsd]{K. Hanold}$^{,2}$ \thanks{\footnotesize Current address: Syagen Technology, 
                                                    1411 Warner Ave., Tustin, CA 92780.},
\author[lbl]{G.J. Wozniak}, and
\author[msu]{D.J. Morrissey}.
\address[osu]{Dept. of Chemistry, Oregon State University, Corvallis,
OR 97331.}
\address[ucsd]{Dept. of Chemistry, Univ. of California, San Diego, La Jolla, CA 92093.}
\address[lbl]{Lawrence Berkeley National Laboratory, Univ. of California, \\
Berkeley, CA 94720. }
\address[msu]{National Superconducting Cyclotron Laboratory, Michigan State\\
University, East Lansing, MI 48824.}


\begin{abstract}
The yields and velocity distributions of heavy residues and fission
fragments from the reaction of 20 MeV/nucleon $^{197}$Au + $^{90}$Zr 
have been measured using the MSU A1200 fragment
separator. A bimodal distribution of residues is observed, 
with one group, resulting from peripheral collisions, having fragment 
mass numbers A=160--200 while the other group,
resulting from ``hard"collisions, has A=120--160.
\ This latter group
of residues can be distinguished from fission fragments by their lower
velocities.
A model combining deep-inelastic transfer and incomplete fusion 
for the primary interaction stage  
and  a statistical evaporation code for the deexcitation stage
has been used to describe the properties of the product 
distributions.
\end{abstract}

\begin{keyword}
Nuclear reactions, incomplete fusion, deep inelastic scattering
\PACS 25.70.-z, 25.70.Jj, 25.70.Lm
\end{keyword}

\end{frontmatter}


\section{Introduction}

The yields of the heavy residues \cite{spectra,sarar}, i.e., the large
remnants of the heavy member of an asymmetric reacting pair of nuclei, are
known to comprise a large fraction of the reaction cross section for
intermediate energy nuclear collisions. However, there are
experimental difficulties in studying the formation of these nuclei in
asymmetric reactions using ``normal'' kinematics. The energies of the
residues are low ($\sim$15 keV/nucleon \cite{spectra,sarar}) 
and the masses are large. 
Experimental thresholds cause one to miss substantial portions
($\>$50\%) of the product distributions \cite{spectra}. 
One can overcome the
problem of detecting low fragment energies by studying these asymmetric
reactions in ``inverse kinematics'', i.e., bombard a low mass nucleus with a
large mass nucleus. By using a high resolution spectrometer/detector system
to observe the projectile-like fragments (PLFs) from such reactions,
one can gain important information about the reaction mechanism, 
complementary to that obtained in exclusive studies,  where only the light
reaction partners are observed with high isotopic resolution.

Pioneering studies by Bazin, et al.\cite{bazin}, Faure-Ramstein, 
et al. \cite{faure}, Pfaff, et al. \cite{pfaff} and Hanold, et al. \cite{karl} 
have  shown the utility of this
approach in studying the heavy reaction products from Kr+X and Xe+X mass
asymmetric collisions at intermediate energies. In a previous paper \cite
{george}, we reported the results of a similar study of the reaction of 20
MeV/nucleon $^{197}$Au with the light nuclei $^{12}$C and $^{27}$Al. 
\ In these cases, we
focussed on {\it fusion-like events} and tried to understand the details of
the particle emission/fission competition in the de-excitation of these hot
nuclei.
\ In this paper, we report the results of an extension of the
previous work to study the yields and velocities of the heavy residues and
fission fragments from the interaction of $^{197}$Au (20 MeV/nucleon) 
with an intermediate mass nucleus, $^{90}$Zr. \ In these systems, fusion-like
collisions are less frequent.
Previous studies \cite{bazin,granier,barz,yokoyama,marchetti,aboufirassi},
\cite{garcia,baldwin,morjean,enterria} of the collisions of heavier nuclei at
intermediate energies have shown binary dissipative collisions to be the
dominant reaction mechanism. \ Evidence has been presented \cite{enterria}
for a sequential decay of one of the initial binary fragments leading to a
three (or more) body final state along with dynamically emitted IMFs from
the collision complex \cite{toke}. \ In some of the binary encounters the
projectile-like fragments have been found \cite{aboufirassi,morjean} to have
very high temperatures (T$\sim $7 MeV).
\ A variety of phenomenological
models have been used, with modest success, to describe these collisions 
\cite{yokoyama,marchetti,garcia}.\ The goal of this extension of previous
work is to understand quantitatively the reaction mechanism(s) responsible
for residue formation in  more symmetric collisions at 20 MeV/nucleon.
We shall present evidence indicating that, in addition to the traditional
evaporation residues from peripheral collisions, some residues are formed in
``hard'' collisions, possibly resulting from  very asymmetric fission and/or 
IMF emission.  The distributions of these  residues, which are very proton-rich,  
cannot be  fully described by available phenomenological models of these 
collisions. 

Although the Au + Zr reaction has not, to the best of our knowledge, been 
studied before, there are a large number of studies 
\cite{otto,raja,dalili1,dalili2,rudolf,dalili3,lucas,solbach,loveland,aleklett},
\cite{zamani,adorno,stuttge,lott,peaslee,skulski1,colin,skulski2,shen} 
of the analogous $^{197}$Au + Kr
reaction at intermediate energies ranging from 4.9 to 70 MeV/nucleon
with some studies \cite{dalili2,dalili3,lucas,zamani} at a beam energy 
of $\sim$20 MeV/nucleon.  What this work adds to this extensive data set is 
detailed, high resolution information on the properties of the heavy, 
Au-like fragments in these reactions garnered from physical rather than 
radiochemical measurements.
(The previous studies of the Kr + Au reaction generally point to the 
presence of both deep inelastic and fusion-like mechanisms.  
The fragments of Au arise as both fission
fragments and evaporation residues \cite{skulski1,skulski2} with the residues
being formed in collisions involving a larger dissipation of the
projectile energy.)

The paper is organized as follows: In Section 2, a brief description of the
experimental apparatus, the measurements and the data analysis is given. In
Section 3, yield distributions are presented and compared with previous
work. Also the velocity distributions are discussed.
In Section 4, the results of the measurements
are compared to modern models of the dissipative features of intermediate
energy collisions.  Finally, conclusions from the present study are
summarized in Section 5.


\section{Experimental}

\subsection{Description of Apparatus}

The experiment was performed at the National Superconducting Cyclotron
Laboratory at Michigan State University  using the A1200 fragment 
separator \cite{brad}. 
This work is a direct extension of the work of Souliotis et al. 
\cite{george}, where the collisions of 20 MeV/nucleon $^{197}$Au with C and Al
were studied. 
It should be noted that use of  the A1200 for very asymmetric collisions
with heavy beams (where the grazing angle is small, i.e. 1-2$^{o}$ 
\cite{Wilcke}), leads to rather efficient collection of the projectile residues 
(that can reach  50\%) at this energy regime.
However,  using the A1200 for more symmetric systems, as for the Au+Zr 
reaction in this  work (where the grazing  angle is 5.5$^{o}$), 
results in substantial loss of  residue cross sections, since the 
measurements are performed near 0$^{o}$, while 
the residue cross sections are expected to peak at larger angles, 
close to the grazing  angle. In this case, only a small fraction of the 
residue cross section can be  collected  (of the order of a few percent).   
Despite this limitation, we attempted to exploit the 
possibility of getting isotopically resolved Au  fragments with this 
apparatus and tried to extract some information regarding the production
mechanism. 
The experimental setup, calibration procedures and data analysis were similar 
as in ref.  \cite{george} and are briefly summarized below.

A 20 MeV/nucleon $^{197}$Au beam, produced by the K1200 cyclotron, 
interacted with a $^{90}$Zr target  of thickness 1.0 mg/cm$^{2}$.
(The target thickness was such that the maximum energy loss of the beam
in traversing the target was 0.26 MeV/nucleon \cite{hubert}.)
The reaction products were analyzed with the  A1200 fragment separator 
operated in the medium acceptance mode [with an angular
acceptance of 0.8 msr ($\Delta \theta =20$ mr, $\Delta \phi =40$ mr) and a
momentum acceptance of 3\%]. The primary beam struck the target at an angle
of 1.0$^{o}$ relative to the optical axis of the spectrometer. 
The A1200 provided two intermediate dispersive images and a
final achromatic image (focal plane). At the focal plane, the fragments were
collected in a three-element ($\Delta $E$_{1}$, $\Delta $E$_{2}$, E) Si
surface barrier detector telescope. The 300 mm$^{2}$ Si detectors were 50,
50 and 300 $\mu $m thick, respectively.

Time of flight was measured between a parallel plate avalanche counter
(PPAC) and a microchannel plate detector (mcp) positioned at the first
dispersive image and at the focal plane, respectively, and separated by a
distance of 14 m. The PPAC at the first dispersive image was also  X--Y 
position sensitive  and  used  to record the position of the reaction
products. The horizontal position, along with NMR
measurements of the A1200 dipole fields, was used to determine the magnetic
rigidity $B\rho$ of the particles.
A series of measurements at overlapping magnetic ridigity settings of the 
spectrometer were performed in the  region  1.500--1.750 
Tesla-meters. This region is below the rididity B$\rho$=1.824 T\,m 
of elastically  scattered  20 MeV/nucleon $^{197}$Au particles at their
equilibrium charge state (Q$_{eq}$=69.5) \cite{baron}.


The reaction products were characterized by an event-by-event measurement 
of dE/dx, E, time of flight, and magnetic rigidity. 
The response of the spectrometer/detector system  to ions of known atomic 
number Z, mass number A, ionic charge q and  velocity was measured  using 
analog beams  as described  previously \cite{george}. \
The procedure to extract  Z, q and A  from the measured quantities is 
outlined below and described in more detail in \cite{george}.

The transmission detectors ($\Delta E_1$ and $\Delta E_2$)  were calibrated 
by directly correlating pulse height with energy loss as calculated using 
the data of Hubert et al. \cite{hubert}.
For the stopping (E) detector, pulse-height defect
(PHD) corrections were necessary to obtain an accurate reconstruction of the
residual energy of the particles. The procedure developed in \cite{george}
was used.  The total
energy E$_{tot}$ of the particles entering the spectrometer (i.e.
the sum of the energies deposited in the Si detectors plus small corrections
due to the presence of the PPACs and mcp ) had an overall resolution (FWHM) of  1.1\%.
The velocity resolution was about 0.5\%.

The determination of the atomic number Z was based on the energy loss of the 
particles in the first $\Delta E$ detector and their velocity and was 
reconstructed using the expression: 

\begin{equation}
Z=a_0(\upsilon )+a_1(\upsilon )\,\upsilon \sqrt{\Delta E}+a_2(\upsilon
)(\upsilon \sqrt{\Delta E})^2  \label{Z_eqn}
\end{equation}

where $\upsilon $ is the velocity of the ion entering the detector and $%
\Delta $E the energy loss. 
In order to determine the functions $a_0(\upsilon )$, $a_1(\upsilon )$ and $%
a_2(\upsilon )$ in the velocity range of interest, we used the data of
Hubert et al. \cite{hubert} \ to obtain the coefficients of Eq. (1) for the Z range
25--85 and in the energy range 12--24 MeV/nucleon by applying a
least-squares fitting procedure at each energy, in steps of 0.5 MeV/nucleon.
The values of each coefficient at the various energies were then
fitted with polynomial functions of velocity. 
The atomic number Z of the
particles was reconstructed from the measured $\Delta $E and $v$ using Eq.
(1) with a resulting resolution (FWHM) of 0.9 Z units for heavy residues
(A=160--200) and
0.6 Z units for fission-like (A=100--140) residues.

The ionic charge $q$ of the particles entering the A1200  was obtained from
the total energy E$_{tot}$,  the velocity and the magnetic rigidity
according to the expression: 
\begin{equation}
q=\frac{3.107}{931.5}\frac{E_{tot}}{B\rho (\gamma -1)}\beta \gamma
\label{q_eqn}
\end{equation}
where E$_{tot}$ is in MeV, B$\rho $ in Tm, $\beta =\upsilon /c$ and $\gamma
=1/(1-\beta ^2)^{\frac 12}$. 
The measurement of the ionic charge q had a resolution  of 0.7 and 0.5 q units 
for heavy  and fission-like residues,  respectively. 
Since the ionic charge must be an integer, we assigned integer
values of q for each event by setting appropriate windows on each peak 
of the q spectrum (see below). 
Using the magnetic rigidity and velocity measurement, the mass-to-charge 
A/q ratio  of each ion was obtained from the expression: 
\begin{equation}
A/q = \frac{B\rho }{3.107\beta \gamma }  \label{Aq_eqn}
\end{equation}
Now, combining the q determination with the A/q measurement, the mass A
was obtained as:
\begin{equation}
A = q_{int} \times A/q  \label{A_eqn}
\end{equation}
(q$_{int}$ is the integer ionic charge determined as above)
with an overall resolution (FWHM) of  1.0 and 0.7 A units, again for 
heavy  and fission-like residues respectively.

The requirement that q be within 
$\pm$0.3 units of its coresponding integer value was applied to 
eliminate contributions of  adjacent q values to the mass assignement.
Similarly, the requirement that Z be within
$\pm$0.3 units of its coresponding integer value was applied to 
avoid  contributions of  adjacent Z  values to the distributions of a 
given element.

Combination and appropriate normalization of the data at the various magnetic
rigidity settings of the spectrometer provided fragment  distributions with respect to 
Z, A, q and velocity. Correction of missing yields caused by charge changing 
at the  PPAC (positioned at the dispersive image) was performed as in \cite{george},
based on the equilibrium charge state prescriptions of Baron et. al. \cite{baron}. 
The distributions were then summed over all values of q. 
It should be pointed out, as mentioned previously,  that the resulting distributions in
Z,A and velocity are residue yield distributions at a  reaction angle of 1$^{o}$ with an 
angular window  of 1.1$^{o}$, in the magnetic rigidity range 1.500--1.750 T\,m.
 

\section{ Results and Discussion}

The extracted  residue yield distributions, sorted by mass number A, atomic number
Z, and velocity $\beta $ ($\equiv$$\upsilon$/c), are shown in 
Figs. 1a, 1b and 1c, respectively.
The yields have been normalized to  beam current 
and target thickness  and are given in mb in Fig. 1a.
The  A distributions are summed over all values of Z and  $\beta$.
The Z--A distributions are summed over all values of $\beta $, and finally,
the $\beta$--A distributions are summed over all values of Z.

It is apparent from examining Figs. 1a and 1b that the spectrometer acceptance 
and resolution have distorted  the fragment yield distributions.  
For example, it was not possible to cleanly
separate the beam nuclei from quasielastic fragments.  Therefore, products that
were very close in mass number, atomic number and energy to the projectile
were either not collected by the spectrometer or rejected in the analysis.  
As a result, a small subset of quasielastic 
events is displayed in Fig. 1.  (The quasielastic events shown in Fig. 1 are very
proton rich relative to the projectile nucleus, which is incorrect.  Their
yields are greatly reduced compared to other fragments which is also incorrect.)
We have chosen therefore to focus on the non-quasielastic events (A $\leq$ 160)
in the analysis.
%
\begin{figure}[tbph]
\vspace{-0.2cm}
\includegraphics[width=9.1cm,height=16.5cm]{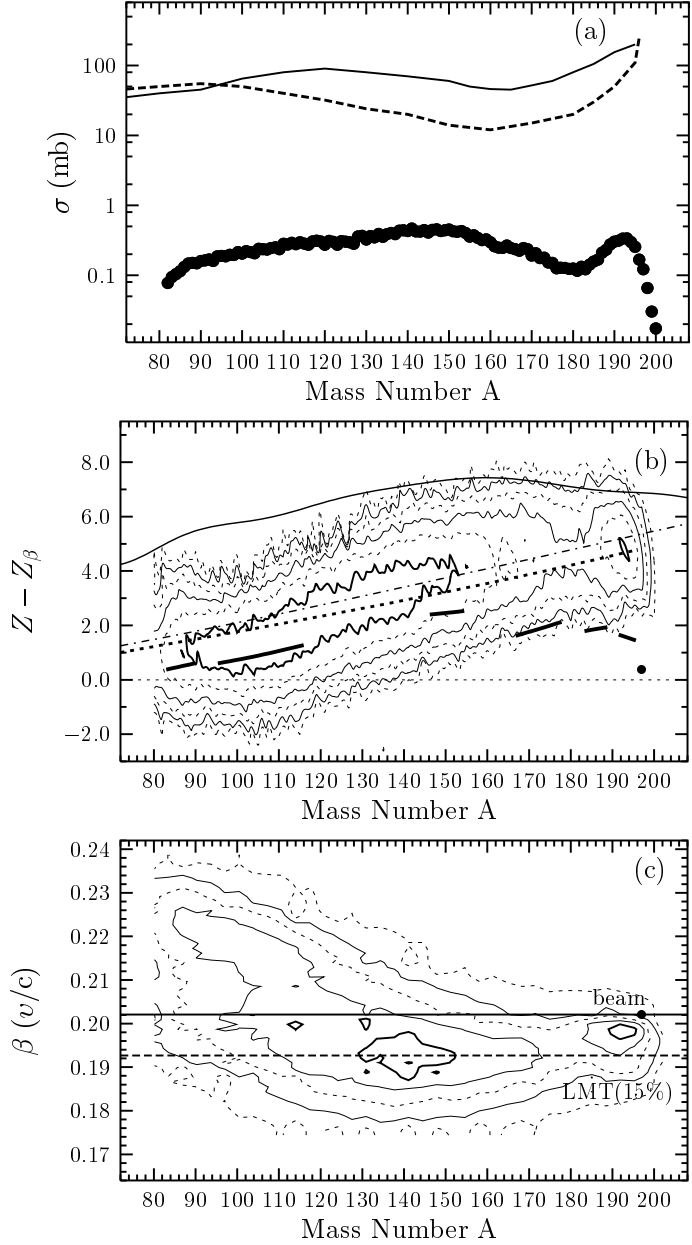}
\centering
\vspace{0.0cm}
\caption{ \footnotesize Fragment distributions for the reaction of 
          20 MeV/nucleon $^{197}$Au  with  $^{90}$Zr.
(a) isobaric yield distribution.
The data are shown as solid points, whereas the solid and dashed 
lines indicate
the yield curves from  radiochemical measurements 
of the reactions 
35 MeV/nucleon $^{84}$Kr+$^{197}$Au \cite{WDL}  and 
21 MeV/nucleon $^{129}$Xe+$^{197}$Au \cite{yokoyama},
respectively.
(b) yield distributions 
as a function of Z (relative to the line of $\protect\beta$ stability,
Z$_{\protect\beta}$) and A.
Highest yield contours are plotted with thicker lines. Successive contours
correspond to a decrease  of the yield by a factor of 2.
The  dashed line indicates the expected values for intermediate energy projectile
fragmentation \protect\cite{pfaff}, the dot-dashed line indicates the position of
the modified ``evaporation residue attractor line''  \protect\cite{bob} and
the solid line indicates the position of the proton dripline
for odd Z nuclei \protect\cite{moller}. Finally, the thick line segments show 
the Z$_p$ values obtained in the radiochemical measurements of 
21 MeV/nucleon $^{129}$Xe+$^{197}$Au \protect\cite{yokoyama}.
(c) velocity vs. mass distributions.
  The definition of the velocity lines is as follows: 
  horizontal full line: beam velocity;  horizontal dashed line: velocity corresponding 
  to 15\% linear momentum transfer  (LMT 15\%).
  }
\label{fig1}
\end{figure}

The mass yield curve (Fig. 1a), has two humps, the distorted quasielastic one 
peaking at A=193  and the other extending from A=170 down to A=80.
The second bump in the mass yield distribution consists of two groups of 
fragments (as can be deduced with the aid of Fig. 1c); one group resulting  from modest 
($\sim$15\%) momentum transfer events with A=120--160 (heavy residues) and another 
group with A=90--120 and  velocities  of forward-moving fission fragments.

 The radiochemical mass yield curves of target-like residues (and fission fragments)
from  the reactions  35 MeV/nucleon $^{84}$Kr+$^{197}$Au \cite{WDL}  and 
21 MeV/nucleon $^{129}$Xe+$^{197}$Au \cite{yokoyama} are also 
shown for comparison (solid and dashed lines, respectively).  
As noted earlier, due to the spectrometer acceptance, the measured yields,
are much lower than the  radiochemical cross sections 
and bear only scant resemblance to them.

In Fig. 1b, the yield distributions of the reaction products are shown as
contour plots of Z versus A. The fragment Z is given relative to the line of
beta-stability, Z$_{\beta }$, which was taken as Z$_{\beta }$ =
A/(1.98 + 0.0155A$^{2/3}$)\cite{marmier}.
The observed fragment distributions lie on the proton-rich side
of stability (Z--Z$_{\beta }=0)$. 
Shown in Fig. 1b, are the centroids of the isobaric charge distributions 
from the radiochemical  measurements of target-like fragments from  the reaction  
21 MeV/nucleon $^{129}$Xe+$^{197}$Au \cite{yokoyama}. The radiochemical centroids 
for  near-Au residues are close to stability due to the dominance of quasielastic 
events in this region. In the symmetric fission fragment region A=90--120, the centroids 
are in fair agreement with the present data, as they both correspond to fission 
fragments. In the region A$\sim$150, the discrepancy may be due to a combination of experimental 
factors (e.g. limited number of measured nuclides in the radiochemical data and acceptance cut 
in the present data). 

Also shown in Fig. 1b are the centroids of the Z distributions seen in intermediate energy
projectile fragmentation \cite{pfaff}.
Despite the fact that no fission is involved in these reactions,   there is some similarity 
(but differences in the slope of Z vs A) between this prescription and the centroids of the 
yield distribution of the present data. 

Charity \cite{bob} has recently discussed the concept of an ``evaporation
residue attractor line''(EAL). \ The EAL is a line in the N--Z plane that
represents the locus of fragment yields produced by the evaporation of
neutrons and charged particles from highly excited nuclei. \ (The
statistical model used to predict the position of the EAL is
GEMINI).
\ In Fig. 1b, we show the position of the modified heavy
fragment EAL for the
decay of highly excited heavy nuclei where the primary fragment is near
the valley of beta stability. \ As shown in this figure, the
centroids of the observed yields are similar to the modified EAL for this
reaction  (A=80--160),  but the slope of Z vs A is not the same.
\ This similarity in charge distributions for the products from
the present reaction and the EAL for A = 120--160 indicates that these 
fragments are  decay products  of highly excited nuclei. 


What causes the centroids of the contours in the data to become more n-rich
compared to these two prescriptions as the fragment mass decreases?  A possible
answer is the occurrence of fission which produces more n-rich fragments than
fragmentation (residue production).  (The two empirical prescriptions
\cite{pfaff,bob} do not include the effects of fission.)  The argument is
that at A=100, a larger fraction of the fragments arise from fission (Fig. 1c)
than at A=150, thus producing the observed slope of Z vs A.  Evidence for this comes
from comparison with the Z distributions measured radiochemically for the
reaction of 20 MeV/nucleon Xe with Au, where the fission of the Au-like nucleus is
more prominent (Fig. 1a,1b).  The observed centroids of the Z distributions for
this reaction are more n-rich than the fragments observed in this work.  (Fig. 1b).
The definitive data comes from comparing Fig. 1b and 1c and noting that
as fission (high velocity events) becomes more prominent, the charge distribution
becomes more n-rich.

One does note that the contours of the observed yields in Fig. 1b lie
close to the position of the proton dripline (plotted for odd Z 
nuclei \cite{moller}). Some of the observed products have 5--7 more protons 
than the stable isobar.  
To examine this point further, we show the detailed fragment charge distributions 
for selected A values for non-quasielastic fragments in Fig. 2.


\begin{figure}[tbph]

\vspace{0.5cm}
\includegraphics[width=11.2cm,height=16.8cm]{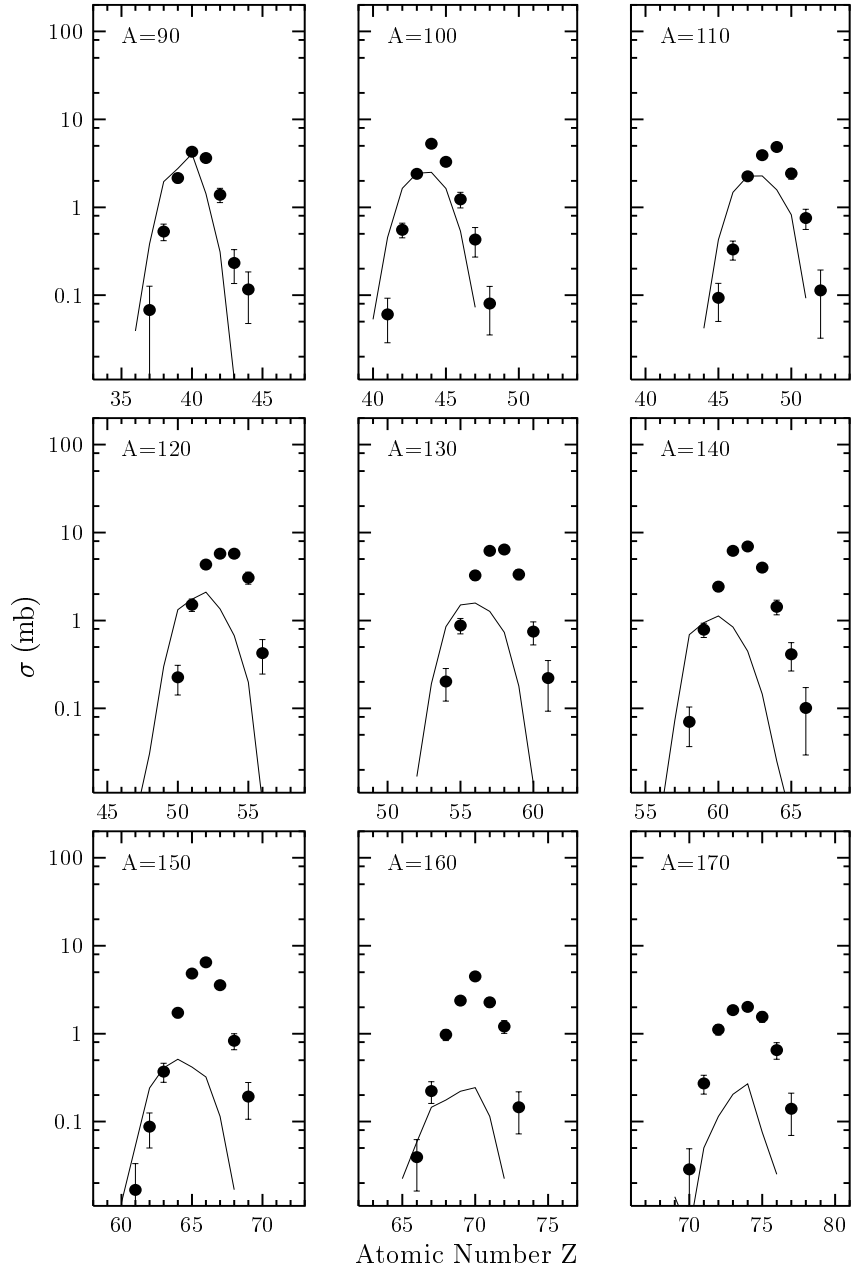}
\centering
\vspace{0.0cm}
\caption{\footnotesize Fragment charge distributions for selected values of the mass
number A for the reaction of 20 MeV/nucleon $^{197}$Au with $^{90}$Zr.
The data are shown as solid points.
The full-line curves are calculations using the DIT+ICF code coupled 
to GEMINI (see text). 
}
\label{fig2}
\end{figure}


For Fig. 2,  an attempt has been made to correct the measured yields for the 
spectrometer angular acceptance, so that we may obtain estimates of fragment production 
cross sections, despite the fact that only a small fraction of the cross section
was collected in the present measurements. Correction factors for the angular 
distributions  of the fission-like fragments (A=90--160) were obtained assuming 
isotropic binary decay of the primary reaction product. For simplicity,  this product 
was assumed to  be a  Au  nucleus with velocity as this observed for the heavy residues 
(which, as already mentioned, correspond to $\sim$15\% linear momentum  transfer) 
and was assumed to scatter off the target at an 
angle somewhat inside the grazing angle. (The grazing angle is 5.5$^o$ degrees 
and the scatter angle was taken to  be 4.0$^o$.) 
Typical correction factors were 40, 60, 90, 150 for masses A=160, 140, 120, 100, 
respectively. 
After applying  mass-dependent correction factors to the measured yields, the 
resulting yields were normalized to values of cross sections of non-isomeric 
independent-yield nuclides from the radiochemical data of the reaction 
(21 MeV/nucleon) $^{129}$Xe+$^{197}$Au \cite{yokoyama}. 
Fragment production cross sections resulted   with  an estimated  
uncertainty of a factor of 2 (due to the correction procedure) and, for several A values, 
are presented  in Fig. 2. 

In examining Fig. 2, we are struck by the large number
of nuclide yields measured for each mass number A, when compared to 
typical radiochemical measurements where, at best, the yields of two or 
three isobars are measured for each A value.
The measured distributions appear to be rather broad. 
The general proton-rich character of these yield distributions indicates that
these intermediate energy collisions may be useful for the
synthesis of heavy  proton-rich nuclei, complementary  to the usual practice 
of synthesizing these nuclei via symmetric compound nuclear reactions involving
the pxn exit channel \cite{cary}.
An attempt to produce and identify proton emitting nuclei using the 
Au+Zr reaction is described in \cite{gas}.

The velocity vs mass distributions (Fig. 1c) show the general absence of   
fusion-like collisions and the prevalence of dissipative collisions. 
As already mentioned, one also sees evidence for the occurrence of fission 
events for A=90--120 with velocities higher than that of the beam, 
corresponding to fission fragments moving forward in the rest frame of the 
fissioning nucleus (the spectrometer settings were such that only 
forward-moving fragments were collected from the two kinematical 
solutions). In addition,  one sees a group of events (A=120--160) with rather constant 
velocity,  lower than that of the beam (by approx. 5\%) corresponding to $\sim$ 15\% 
fractional linear momentum transfer (FLMT). 
The existence of a constant fragment velocity,
roughly independent of fragment mass, is not consistent with fission, 
where the fragment velocity should be inversely correlated with the fragment 
mass (as it is for A=90--120 for the Au-like fragment) \cite{dalili}. 
We associate this group  of fragments (A=120--160) mainly with 
residues that have emitted one or more IMFs in accord with  
a recent radiochemical study of  target-like residues from the
$^{208}$Pb (29 MeV/nucleon) + $^{197}$Au reaction  \cite{pbau}. 
From the velocity of this group of fragments (relative to that of the beam) we can estimate 
an average total kinetic energy loss of $\sim$350 MeV.

We also note in Fig. 1c,   that the fragment group A=90--120, along with the fission fragments, 
contains  residues characterized by  constant velocity (as in the A=120--160 group). 
It is rather remarkable that residues with less than half the mass of the projectile  nucleus 
are observed in this reaction.


\section{Comparison with Reaction Simulations}

To gain insight into the underlying reaction mechanisms, we have performed 
simulations  using an appropriate phenomenological model.  The basic features
 of the reaction mechanism model are pre-equilibrium emission, dissipative
 deep inelastic transfer (DIT) and for the most central collisions, an incomplete
 fusion component (ICF).  The hybrid model we used that included these features
 has been described by  Veselsky \cite{martin2}.

For each event corresponding to a given impact parameter (partial wave), the effect of
pre-equilibrium emission was calculated using a variant of the exciton
model. Then the interaction of the projectile
and target nuclei are simulated using the Tassan-Got/Stephan model \cite{tassan}
for deep inelastic transfer assuming stochastic nucleon exchange (in the region of 
orbital angular momentum
$\ell$ = 300--870).  For trajectories where the overlap between projectile
and target nuclei exceeds 3 fm, it is  assumed that incomplete fusion takes place 
with this
mechanism being modeled using a variant of the abrasion-ablation model \cite{gosset}.
Following the creation of the primary fragments by this hybrid mechanism,
the statistical de-excitation of the excited primary fragments was simulated
using GEMINI \cite{charity}.  The results of this calculation were then filtered
by  the angular and momentum acceptance of the spectrometer.  This model
has been used successfully \cite{martin2}
to describe the residues formed in the 35 MeV/nucleon Kr + Au reaction 
\cite{skulski1,skulski2} and the 20 MeV/nucleon Au + Ti reaction 
studied by us \cite{AuTi}.

Simulations were run for partial waves from $\ell$=0--870.  De-excitation of the
primary fragment distribution was done using GEMINI.  
This statistical deexcitation code uses Monte Carlo techniques
and the Hauser-Feshbach formalism to calculate the probabilities for 
fragment emission with Z$\leq$2. Heavier fragment and fission fragment 
probabilities are  calculated using the transition state formalism of 
Moretto \cite{moretto}.
In the GEMINI calculations, we have used Lestone's temperature dependent 
level density parameter \cite{lestone}, a fading of shell corrections with
excitation energy and we enabled IMF emission. Finally, fission delay was 
enabled  with parameters as in \cite{george}.  Each partial-wave distribution
was appropriately weighted 
and combined to give the overall fragment
A, Z  and velocity distributions (Figs. 3 and 4).


\begin{figure}[tbp]

\vspace{0.5cm}
\includegraphics[width=9.1cm,height=11.2cm]{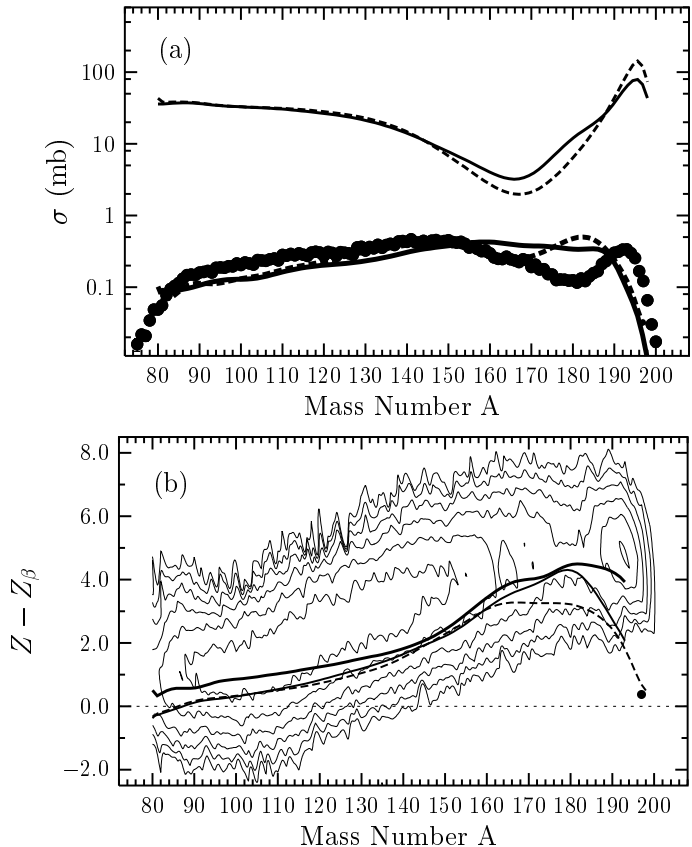}
\centering
\vspace{0.0cm}

\caption{\footnotesize Comparison of measured and calculated fragment 
distributions for the reaction of 20 MeV/nucleon $^{197}$Au with
$^{90}$Zr.
(a) isobaric yield distributions. The data are shown as solid points. 
    The upper full line is the result of DIT+ICF/GEMINI calculation and the 
    dashed line is  from DIT/GEMINI (see text). The lower full line and 
    dashed lines  are the results of the  same calculations as the upper ones, 
    but with a cut corresponding  to the angular and momentum acceptance 
    of the spectrometer.  
(b) charge distributions.  The data are shown as contours;
    the calculated values from DIT+ICF/GEMINI are shown as i) thick full line:
 with acceptance cut and, ii) thin full line: without acceptance cut, 
and the ones from DIT/GEMINI are shown as a dashed line (without acceptance cut). 
}
\label{fig3}
\end{figure}


\begin{figure}[tbph]

\vspace{0.5cm}
\includegraphics[width=9.1cm,height=11.2cm]{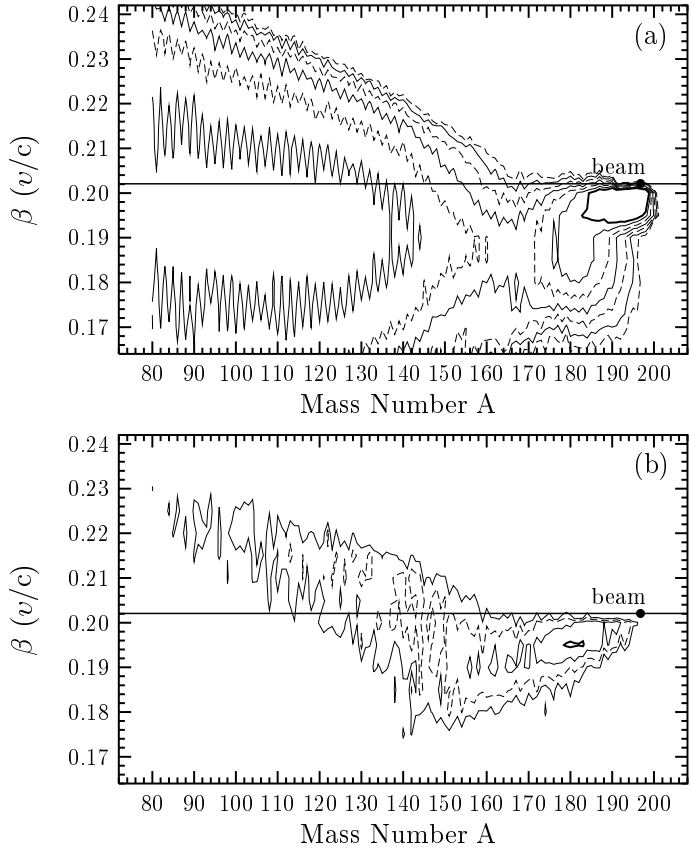}
\centering
\vspace{0.0cm}
\caption{\footnotesize The calculated (DIT+ICF/GEMINI) fragment velocity distributions 
                       (a) without and (b)
                        with the spectrometer acceptance cut included in the calculation.
}
\label{fig4}
\end{figure}


\begin{figure}[tbph]

\vspace{0.5cm}
\includegraphics[width=8.4cm,height=12.6cm]{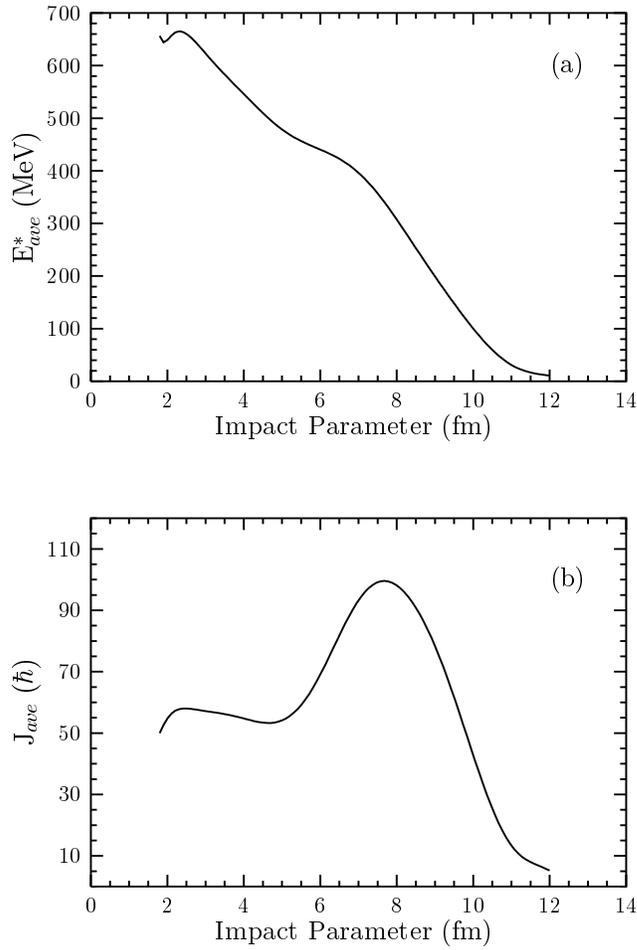}
\centering
\vspace{0.0cm}
\caption{\footnotesize The calculated (DIT+ICF/GEMINI) primary fragment mean excitation 
         energy (a)
         and angular momenta (b) as a function of impact parameter
         for the reaction of 20 MeV/nucleon $^{197}$Au with $^{90}$Zr.
       }
\label{fig5}
\end{figure}


\begin{figure}[tbph]

\vspace{0.5cm}
\includegraphics[width=9.1cm,height=7.0cm]{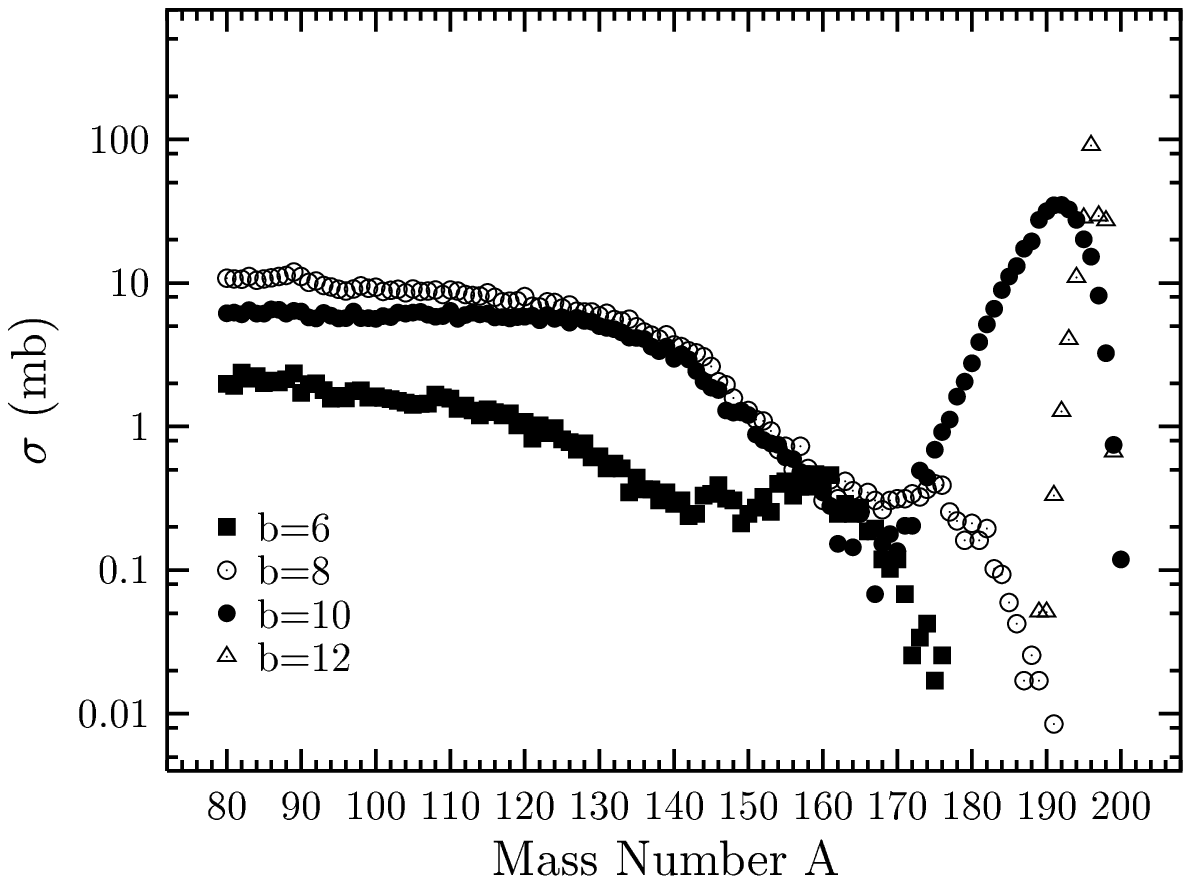}
\centering
\vspace{0.0cm}
\caption{\footnotesize The calculated (DIT+ICF/GEMINI) fragment mass distributions 
corresponding to
selected values of the impact parameter b (in fm) for the reaction of 20 MeV/nucleon 
$^{197}$Au with $^{90}$Zr. 
}
\label{fig6}
\end{figure}


In Fig. 3a,  we compare the measured and calculated mass yield curves for a
number  of conditions.  
The calculations were performed neglecting small
impact parameters (deep inelastic transfer only, DIT/GEMINI) or including
all impact parameters (DIT+ICF/GEMINI).
The upper dashed line is from  DIT/GEMINI calculation and the 
upper solid line is from  DIT+ICF/GEMINI calculation (without filtering 
with  the spectrometer acceptance). The corresponding lower curves are from
the same calculations in which filtering with  the spectrometer acceptance has 
been peroformed.
As we see, the shape of the mass yield curve is
significantly altered by the overall spectrometer cut. 
Note however, the similarity of shapes  between the measured yield curve
and the calculated yield curves filtered with  the spectrometer acceptance.

In Fig. 3b, the measured Z vs A yield distribution is compared with the 
predicted positions of Z$_{p}$, the most probable fragment atomic number for a 
given A value (given 
with respect to the line of $\beta$ stability). Displayed are the results of 
DIT/GEMINI (dashed line,  no acceptance filter), and of DIT+ICF/GEMINI 
(no acceptance filter: thin solid line, with spectrometer acceptance filter: thick line). 
We observe that the latter calculation with the acceptance cut does a fair job in 
describing the 
Z$_p$ values of the A=160--190 residues, but predicts more n-rich fragments, 
consistent with fission, 
in the region A$<$160. Thus, the  heavy residue group A=120--160 seen in the data is not 
predicted by the calculation,  due to the prevalence of fission as a deexcitation channel
in the GEMINI calculation.

The calculated fragment charge distributions are compared with the data in Fig. 2.  
The calculated  centroids of the distributions are too n-rich as noted above,
but the predicted widths of the charge distributions agree  rather well with the 
observations.

 In Fig. 4, we show the predicted velocity distributions (which are to be
compared to Fig. 1c). In Fig. 4a, we show the unfiltered velocity
distribution, while in Fig. 4b we show the velocity distribution as filtered
by the spectrometer angular and momentum acceptance.  Qualitatively, 
some  features of the measured distributions are reproduced, i.e., 
the appearance of residue events
(A = 160--190) with mean velocities similar to the observed velocities and
the characteristic forward moving fission fragment velocity pattern for the lighter
fragments (A$<$120) which are also similar to the observations. 
However, as already pointed out, the predicted velocities of the A=120--160 
group are inversely correlated with mass and (along with their 
average Z values) are consistent with fission, whereas 
the measured properties of this  fragment group are those of residues.

We conclude that the hybrid model proposed  by Veselsky, 
coupled to a modern statistical de-excitation code, 
can describe some gross features of the  residue distributions, especially
those not too far from the projectile nucleus, whereas for lower mass
residues it predicts mostly fission.
We might then ask what we might learn about residue formation from the
simulation model only.
To gain  insight  regarding the primary interaction stage, 
we show in Fig. 5 the simulated
properties (DIT+ICF/GEMINI) of the collision as a function of 
impact parameter b.
Very large  excitation energies and angular momenta are involved 
in these collisions.
Interestingly, the average spin of products in the impact parameter
range 6--8.5 fm is predicted to be higher than the maximum angular momentum 
for which the symmetric fission barrier vanishes (approx. 80$\hbar$) which causes these
nuclei to fission.

In Fig. 6, we show the simulated fragment mass distributions arising from 
collisions with various impact parameters. 
From these impact parameter sorted mass
distributions, we see that the near projectile fragments originate 
in very peripheral
collisions (quasielastic events) with (Fig. 5) low associated excitation
energies and angular momenta. The rest of the fragments are produced by
collisions involving a large range of impact parameters (and thus the
fragment mass number A cannot be used as an impact parameter trigger).
These broad mass distributions represent events where the heavy fragment
from the
collision either fissioned symmetrically (A=90--120) or fissioned very 
asymmetrically or emitted one or more IMFs in its de-excitation (A=120--160).
As already mentioned, the measured velocity spectra (Fig. 1c) indicate 
that these fragments (A=120--160) are not the result of symmetric fission. 


\section{Conclusions}

What have we learned from this high resolution experimental study of the
residues produced in the reaction 20 MeV/nucleon $^{197}$Au + $^{90}$Zr ?  Two findings
seem of particular importance.  They are: (a) the observation of surviving
residues (A=120--160) which appear to result from ``hard collisions''.
They are relatively neutron-deficient with broad charge distributions and 
velocity, on average, corresponding to a linear momentum transfer of $\sim$ 15\%.
(b)  A semi-quantitative
description of heavy residue formation in intermediate energy collisions between
massive nuclei by the hybrid model of Veselsky followed by the statistical deexcitation 
code GEMINI was found.
However, the calculation was not able to describe the residue group with 
A=120--160, and gave primarily fission fragments in the whole mass range 
A$<$160.

We gratefully acknowledge the support of the A1200 group and operations
staff of Michigan State University during the measurements and the help of
L. Hart in the analysis of the data. One of us (G.A.S) is also thankful to 
M. Veselsky for his help in the course of calculations and stimulating  
discussions.
Financial support for this work was given, in part, by the U.S. Department 
of Energy under Grant No.
DE-FG06-88ER40402, No. DE-FG03-97ER41026, and Contract DE-AC03-76SF00098
and the National Science Foundation under Grant No. PHY-95-28844.




\end{document}